\documentclass{article}
\usepackage{spconf,amsmath,graphicx,hyperref}

\usepackage[utf8]{inputenc} 
\usepackage[T1]{fontenc}    
\usepackage{url}            
\usepackage{booktabs}       
\usepackage{amsfonts}       
\usepackage{nicefrac}       
\usepackage{microtype}      
\usepackage{xcolor}         
\usepackage{colortbl}
\usepackage{multirow}
\usepackage{tabularx}
\usepackage{subcaption}

\definecolor{Gray}{gray}{0.9}
\newcolumntype{a}{>{\columncolor{Gray}}c}
\newcolumntype{b}{X}
\newcolumntype{d}{>{\hsize=.13\hsize}X}
\newcolumntype{s}{>{\hsize=.37\hsize}X}
\newcolumntype{x}{>{\centering}>{\hsize=.13\hsize}X}
\newcolumntype{e}{>{\centering}c}
\newcolumntype{f}{>{\columncolor{Gray}}>{\centering}c}
\newcolumntype{g}{>{\arraybackslash}>{\columncolor{Gray}}>{\centering}c}
\arraybackslash

\title{On Deepfake Voice Detection - It's All in the Presentation}

\name{
\begin{tabular}{@{}c@{}}
Héctor Delgado, Giorgio Ramondetti, Emanuele Dalmasso, \\ Gennady Karvitsky, Daniele Colibro, Haydar Talib
\end{tabular}
}

\address{Microsoft}

\begin{document}
\ninept
\maketitle

\begin{abstract}
  While the technologies empowering malicious audio deepfakes have dramatically evolved in recent years due to generative AI advances, the same cannot be said of global research into spoofing (deepfake) countermeasures. This paper highlights how current deepfake datasets and research methodologies led to systems that failed to generalize to real world application. The main reason is due to the difference between raw deepfake audio, and deepfake audio that has been \textit{presented} through a communication channel, e.g. by phone. We propose a new framework for data creation and research methodology, allowing for the development of spoofing countermeasures that would be more effective in real-world scenarios. By following the guidelines outlined here we improved deepfake detection accuracy by 39\% in more robust and realistic lab setups, and by 57\% on a real-world benchmark. We also demonstrate how improvement in datasets would have a bigger impact on deepfake detection accuracy than the choice of larger SOTA models would over smaller models; that is, it would be more important for the scientific community to make greater investment on comprehensive data collection programs than to simply train larger models with higher computational demands.
\end{abstract}

\section{Introduction}
With the advances in generative AI in recent years, it is increasingly difficult to discern if content is human- or machine-generated. In the case of voice, this realism is achieved using voice conversion or Text-to-Speech (TTS) \cite{tan2021surveyneuralspeechsynthesis} technology. Current generation technology can be used to create voice clones of specific individuals using just a few seconds of speech, and in fact the human ear can no longer differentiate real voices from AI-generated ones \cite{DIEL2024humansvsdeepfakes,groh2024humandetectionpoliticalspeech}. 

While there are legitimate applications of such synthetic voices, they can also be used maliciously in the form of audio \textit{deepfakes}. Examples of misuse are to attribute false statements to a person for reputational damage or spreading misinformation, or defrauding individuals or organizations for large sums of money \cite{chen2024cfodeepfake}.

Research and development of \textit{spoofing countermeasure} (or deepfake detection) systems has carried on for decades. Recently it began to emerge, however, that the majority of datasets created for deepfake detection research are so unrealistic that they were nearly solved via shortcut learning of features that are irrelevant to the spoofing problem, let alone real-world application \cite{asvspoof2021,muller21_asvspoof}.

\begin{figure}
    \centering
    \includegraphics[width=1.0\linewidth]{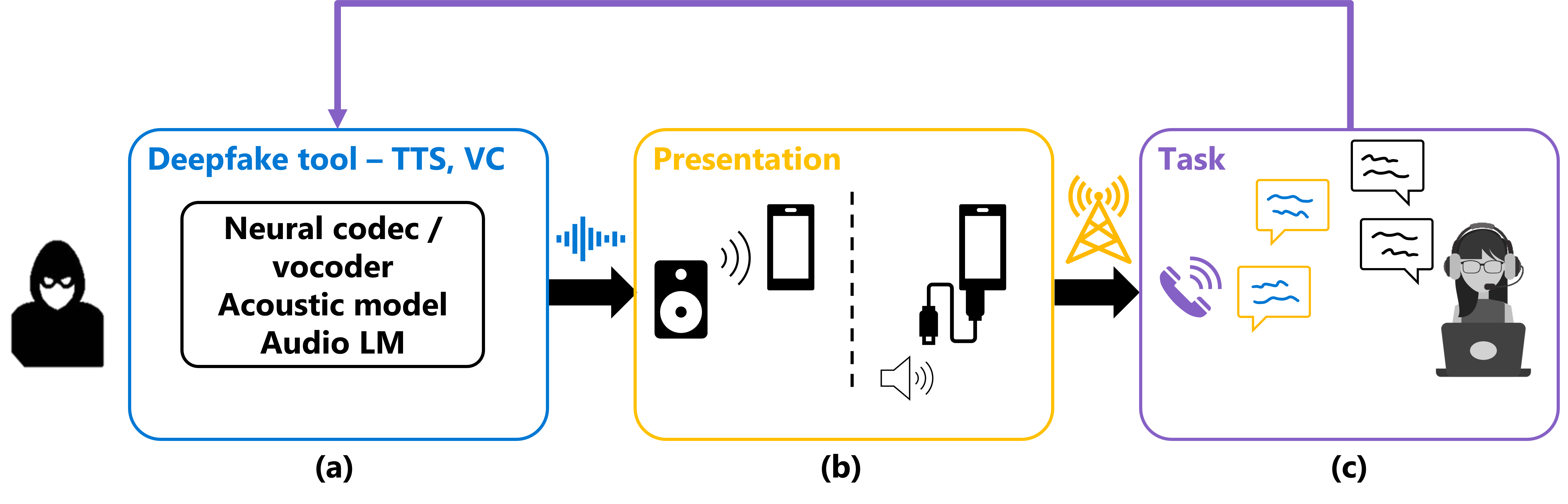}
    \caption{The sequence of steps transforming the voice signal is more elaborate in real-world application (telephone banking call center in this example) than simply using the source deepfake audio file created by the TTS tool (a). In (b), the \textit{presentation} phase, the fraudster presents the deepfake voice through the phone using a loudspeaker or via direct injection technique, which is then transmitted through the telephony network. In the task phase, we arrive at the real-world setting - a phone call takes place to engage the bank's call center agent (c). The feedback arrow indicates that the fraudster uses the deepfake tool with the end task in mind, e.g. by creating conversational phrases. Each phase of this sequence introduces one or more distortions to the source deepfake audio signal.}
    \vspace{-10pt}
    \label{fig:spoofattack}
\end{figure}

We argue that there is a much larger problem in the benchmark deepfake datasets used by the scientific community: the lack of realism. Figure \ref{fig:spoofattack} illustrates the key phases of what we can call the spoof attack sequence, of which deepfake audio generation is merely a first (albeit important) step (phase (a) in the figure).

The majority of current spoofing countermeasure datasets and publications only capture the first phase of the spoof attack pipeline. In this initial phase, audio clips are pristine, and unaffected by signal distortion or other environmental factors. Furthermore, they are sometimes produced using studio-quality recordings of participants. While we now begin to see some progress towards realism \cite{asvspoof5_db,muller25_interspeech}, so far there have only been analyses on individual characteristics (e.g. silences, loudspeaker playback) rather than a more global view. 

In this work, we introduce the first holistic view of a realistic deepfake attack scenario for the purposes of creating richer datasets that will more successfully translate to the real world. We do this by being the first to incorporate direct-injected and loudspeaker playback presentations of deepfakes as well as the dynamics of live conversation into a unified framework, which also includes establishing a broader evaluation benchmark than previously seen.

We will also demonstrate how, in comparing several model scales of SOTA systems, the need for improved dataset creation is more important than the pursuit of ever-increasing model size and computational cost (though they do help, of course).

\begin{table*}[t]
    \centering
    \caption{Details on training and test datasets. \textbf{Base:} public dataset or raw data we created. \textbf{Presented:} loudspeaker playback or direct injection of deepfake audio into a phone call. \textbf{Augmented:} Augmented data from neural codecs and vocoders. \textbf{Realworld:} Fraud Academy dataset.}
    \resizebox{\textwidth}{!}{\begin{tabularx}{1.2\linewidth}{dsxxb}
    \toprule
    \multicolumn{5}{c}{\textbf{TRAINING DATA}}  \\
    \toprule    
    Category	&	Dataset 	&	\#Deepfake	&	\#Bonafide	&	Comments \\
    \toprule 
    \multirow{ 4}{*}{Base}	&	ASVspoof 2019 LA~\cite{wang20_asvspoof}	&	22,800	&	2,580	&	Based on VCTK. Official training set. 6 generation algorithms. \\
    	&	ASVspoof 5~\cite{asvspoof5_db}	&	163,560	&	18,797	&	Based on MLS English. 8 generation algorithms. Raw, wideband data. \\
    	&	SWB-Synth/raw	&	4,500	&	1,000	&	Switchboard voices cloned with ElevenLabs, Play.ht and OpenAI. \\
    	&	MLS-Synth/raw	&	7,979	&	1,989	&	MLS English voices cloned with ElevenLabs, Mars5, YourTTS, Play.ht and OpenAI. \\
    \midrule
    \multirow{ 4}{*}{Presented}	&	SWB-Synth/inj.	&	13,500	& - &	`SWB-Synth/raw' data injected into phone call.	\\
    	&	SWB-Synth/play.	&	10,500	& - &	`SWB-Synth/raw' data played back through a loudspeaker into
phone call.	\\
    	&	MLS-Synth/inj.	&	23,937	&	5,965	&	`MLS-Synth/raw' data injected into phone call. Bonafide data injected to apply telephone processing. \\
    	&	MLS-Synth/play.	&	15,958	& - &	`MLS-Synth/raw' data played back through a loudspeaker into
phone call.	\\
    \midrule
    Augmented	&	neural vocoders / codecs augmentation	&	213,785	&	263,770	& Public datasets (VoxCeleb 1\&2, Fisher Spanish, Mixer6, Mixer Phone, RSR2015, SITW, SRE 16/18/19, Switchboard) processed with neural codecs and vocoders. \\
    \toprule
    \multicolumn{5}{c}{\textbf{TEST DATA}}  \\
    \toprule    
     \multirow{ 8}{*}{Base}		&	ASVspoof 2019 LA	&	63,757	&	7,335	&	Official test set. 13 generation methods. \\
    		&	ASVspoof 2021 LA~\cite{asvspoof2021}	&	133,096	&	14,771	&	Official test set. 13 generation methods. Raw and speech/general-purpose codecs. \\
    		&	ASVspoof 2021 LA Hidden &	14,853	&	1,949	&	Official `Hidden Track' test split where silences are trimmed from the audio segments. \\
    		&	ASVspoof 2021 DF~\cite{asvspoof2021}	&	517,603	&	14,837	&	Official test set. $>$100 generation methods. Raw and general-purpose codecs. \\
    		&	ASVspoof 5 w/o Encodec	&	466,350	&	119,377	&	Official test set, excluded C4 and C7 due to Encodec use in training data. Based on MLS English. 16 generation algorithms. Varied narrowband and wideband codecs. Includes adversarial attacks.\\
    
    		&	In-the-wild~\cite{inthewild}	&	11,539	&	18,750	&	Full dataset. Audio from YouTube videos.\\
    		&	SpoofCeleb~\cite{spoofceleb}	&	81,224	&	9,103	&	Official test set. Based on VoxCeleb1. 9 generation algorithms. Raw synthetic data.\\
    
        &	Pool	&	21,000	&	21,000	&	3K deepfake and 3K bonafide randomly sampled from each test dataset. \\
    \midrule

    \multirow{ 2}{*}{Realworld}	&	Injected TTS	&	295	&	\multirow{ 2}{*}{1,143}	&	\multirow{ 2}{*}{\shortstack[l]{Self-collected DB featuring real authentic and fraud phone calls }} \\
    	&	Playback TTS	&	825	&		&	 \\
    \bottomrule

    \end{tabularx}}
    \label{tab:data}
\end{table*}

\section{A Guide to Realistic Deepfake Data}
\label{sec:data}
To achieve the most effective performance of a deepfake detection system for the telephone banking scenario, the dataset creation process must go through the full process illustrated in Figure~\ref{fig:spoofattack}. Fraudsters would have sourced original recordings of their target and used these to generate deepfake audio clips, speaking phrases that will allow for financial crime (``I'd like to transfer some money''). The original deepfake recordings are then distorted by one or more factors through each phase in the sequence.

In this subsection we describe the different types of treatments used to develop the deepfake detection training and test datasets, with details appearing in Table~\ref{tab:data}. Data categories 2 \& 3 are of particular note here, as they introduce the effects of \textit{presentation} and \textit{task}, which have not previously been explored in deepfake detection research or data collection. We used a voice activity detection (VAD) module~\cite{nuance_vad} during the creation of all datasets to omit silences, as one way to help mitigate the impact of superficial signal manipulation.

\textbf{Data Category 1 (Base) - Public datasets and generation tools} consists of existing public antispoofing datasets, and new deepfake datasets we created using two public datasets as seeds (Switchboard~\cite{switchboard} and the English part of Multilingual LibriSpeech (MLS)~\cite{mls}), and Eleven Labs, play.ht, OpenAI Voice Engine~\cite{voiceengine}, Mars5~\cite{mars5repo}, and YourTTS~\cite{casanova2022yourtts} as TTS engines. The key difference here compared to other approaches to training deepfake detection models is that we combined datasets from multiple sources, rather than focus on specialized subsets; the idea is to favor models that will more effectively generalize. Although datasets in this category lack realism, in aggregate they still provide rich material on which to train deepfake detection models.

This group contains  $\sim$1.7 million audio samples (average 4.4s net speech duration), and is used for both training and testing.

\textbf{Data Category 2 (Presented) - \textit{Presented} audio datasets} was built by processing raw data (as returned by TTS engines) through the presentation phase (Figure~\ref{fig:spoofattack}(b)), however using only a subset of the data described in the \textit{Base} category. For the deepfake audio presentation, we used two different smartphones as calling devices, namely the Samsung Galaxy A12 and Redmi Note 8 Pro. For direct audio injection, we used two methods: digital injection through Bluetooth, and analog injection through a wired connection into the smartphone’s microphone input. As playback devices, we used an ESI Aktiv studio monitor and JBL Charge Essential portable Bluetooth loudspeaker. All the recordings were done in a room with no special acoustic treatment. The entire process was automated, with calls going into a staged call center for this purpose. This yielded $\sim$70 thousand audio samples, which were only used for model training.

\textbf{Data Category 3 (Realworld) - Fraud Academy} is a private dataset that was developed in a live collection with 80 participants. This dataset was intended to recreate real world conditions, consisting of all the phases in the fraud attack sequence (Figure~\ref{fig:spoofattack}). Participants were provided scripts and asked to otherwise improvise conversations with call center agents as they role-played between legitimate and fraudulent calls, leveraging state of the art TTS tools of their choice to create deepfake voices. Participants spanned multiple demographic groups, used a variety of devices, and were located in multiple locations across the world, allowing for a richness of characteristics in the data. As a result there were at least 16 calling device types (from 5 manufacturers, e.g. Apple, Samsung) and 5 loudspeaker types used across the 80 participants, as well as 10 TTS engines.

This collection resulted in a total of 2,263 phone calls segments, of which 1,120 contained spoof attacks (either injected or replayed). The recordings were segmented into clips of 20 seconds of net speech. Despite its small size, to our knowledge this dataset still represents the most realistic and diversified dataset that has been studied so far.

The Fraud Academy dataset was completely held out from all deepfake detection model training and development steps, allowing us to test a model's ability to generalize and real-world effectiveness.

\textbf{Data Category 4 (Augmented) - vocoder-synthesized data} is only used to augment training datasets, done by processing bonafide human speech using neural vocoder~\cite{wang23vocoders} and codec~\cite{xie2024codecfake} augmentation. i.e. ``pseudo-spoof'' waveform data was created at scale by using the same components at the core of TTS algorithms. We used the HIFI-GAN~\cite{hifigan}, WaveGrad~\cite{wavegrad} and WaveNet~\cite{wavenet} vocoders, as well as the Encodec~\cite{defossez2022highfi} and Vocos~\cite{siuzdak2023vocos} codecs.

\section{Evaluation}
\label{sec:expresults}

In this section we demonstrate quantitatively how the deepfake dataset creation methods described in section~\ref{sec:data} are necessary for developing deepfake detection systems that generalize to real-world application, and that this is more impactful to accuracy than brute force approaches such as increasing model size and computation cost.

\subsection{Deepfake detection systems}
\label{sec:system_description}

We evaluated three SOTA systems that also differ by frontend approach; we used either general-purpose acoustic features (log-mel filterbank coefficients) or a large self-supervised learning (SSL) speech representation model (pre-trained WavLM~\cite{wavLM} Large\footnote{https://huggingface.co/microsoft/wavlm-large} with 316.62M parameters, kept frozen during the backend training).

\begin{figure}[]

\begin{minipage}[b]{1.0\linewidth}
  \centering
  \centerline{\includegraphics[width=\linewidth]{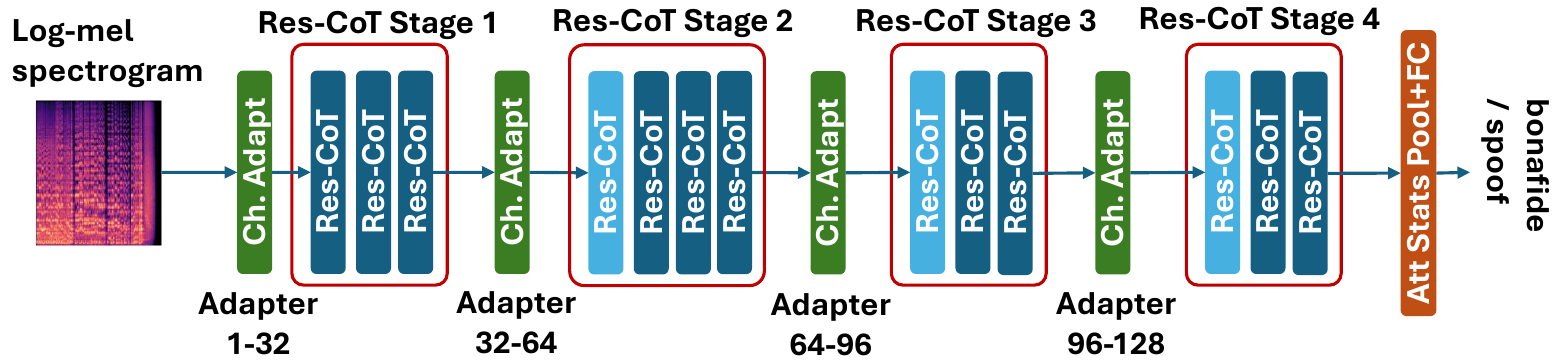}}
  \centerline{(a)}\medskip
\end{minipage}

\begin{minipage}[b]{1\linewidth}
  \centering
  \centerline{\includegraphics[width=0.4\linewidth]{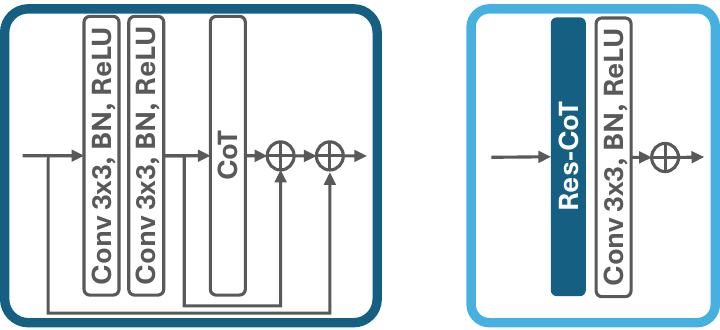}}

  \centerline{(b)}\medskip
\end{minipage}

\caption{(a) The ResNet-CoT system for spoof detection and (b) The two variants of Res-CoT blocks used by the model (encoded by dark and light blue colors in (a)).}
\label{fig:coatnet}

\end{figure}

\textbf{logmel-ResNet-CoT} (Figure~\ref{fig:coatnet}(a)) uses acoustic features (64 log-mel-band spectrograms) and is a Residual Network with Contextual Transformers~\cite{li2023coatnet}, a transformer-like 2D self-attention mechanism originally designed for visual recognition, but modified and applied for audio here, for the first time. The self-attention mechanism (CoT) is incorporated to the end of each residual (Res-CoT) block, as shown in Figure~\ref{fig:coatnet}(b). Four Res-CoT stages are each preceded by adapters (2D-Convolutional layer, batch normalization and ReLu) which increase the number of channels, and are followed by an attentive statistics pooling~\cite{nec_attentive_pooling} layer and a fully-connected (FC) layer. The model has 3.55M parameters.

\textbf{WavLM-LLGF} is a variant of the system described in~\cite{wang2022investigatingselfsupervisedendsspeech} which uses an SSL frontend that feeds an LLGF backend composed of a light convolutional neural network (LCNN) followed by two bi-directional long short-term memory (LSTM) units, a global average pooling layer, and a FC output layer. Unlike the original system, which uses the SSL frontend's last transformer layer output, we aggregate the intermediate outputs of the CNN feature encoder and each 24 transformer layers of a pretrained WavLM model as follows: projection to 128-dimensional space $\rightarrow$ GeLU$ \rightarrow$ LN $\rightarrow$ weighted sum $\rightarrow$ LN, where LN indicates layer normalization and GeLU indicates Gaussian error linear unit. Including the WavLM frontend, the model has 317.70M parameters

\textbf{WavLM-Nes2Net} is described in~\cite{liu2025nes2net}. It uses an SSL frontend which feeds a Nested Res2Net (Nes2Net) backend: the intermediate outputs of a pretrained WavLM model are fed through a feature aggregation layer and form the input to an outer layer, in which the high-dimensional features are uniformly split into subsets. Each subset, except for the first, is fed to a corresponding nested Res2Net module. The nested module output is then both concatenated with outputs from the other subsets and added to the next subset, before the next nested module. The output goes through temporal pooling and a FC layer. Including the WavLM frontend, the model has 316.93M parameters.

All systems use a classification head with two outputs (one per class). In inference, the detection score $s$ is obtained by combining the output logits $l_{spoof}$ and $l_{bonafide}$ as $s=0.5(l_{spoof}-l_{bonafide}$). Higher (lower) scores favor the spoof (bonafide) hypothesis.

\subsection{Experimental setup}
\label{sec:experimental_setup}

\textbf{Training conditions:} Training sets were formed by combining different permutations of the categories described in Section~\ref{sec:data} (Realworld was held out, only used as a test set). Specifically, we trained models using the following combinations: Base, Base+Augmented, Base+Presented and Base+Presented+Augmented. Recall that Base is most similar to common lab setups in literature.

\textbf{Test conditions:} In reporting average accuracy across the Base datasets, we favor the pooled assessment, which suggests one decision threshold is used across all test sets. This will be one more measure of how well a given model generalizes across attacks, even unseen ones. The final Base test set was composed of randomly selected 3k bonafide and 3k spoof segments from each of the original test sets (Table~\ref{tab:data}). For the Realworld test set, we report results for injection and playback presentation methods separately.

\textbf{Metrics:} A key selection criterion for any candidate deepfake detection model will be its accuracy; this is often presented as Equal Error Rate (EER), which we report here for the sake of easy comparison to literature. The more meaningful measure will be the Missed Detection Rate (MDR) at a target False Alarm Rate (FAR), which we will set to 1\%. Metrics are calculated using the full audio lengths for Base, whereas for Realworld, results are obtained by averaging the performance across 6 different decision \textit{checkpoints} (using 2, 3, 6, 9, 12 and 15 seconds of net speech).

\textbf{Implementation details:}
We use various online data augmentation methods during training, including speech/audio codec processing, volume variation, and RawBoost~\cite{rawboost}). For all experiments and models, audio is sampled at 8 kHz and a voice activity detection module (VAD) is used to drop non-speech frames before feeding the models. The DNNs are trained on variable length batches (random crops in the range of 0.9-1.2s and 1.8-2.4s net speech for the logmel-ResNet-CoT and wavLM-based systems, respectively) of 256 examples, with a cross-entropy loss and the AdamW optimizer for 780K and 29K iterations (logmel-ResNet-CoT and wavLM-based systems, respectively). To train each model, we used a node equipped with 96 AMD EPYC 7V12 vCPU, 1800 GiB memory, and 8 NVIDIA A100 GPUs (from which only 2 are used per model). During testing, we discarded every segment with less than 0.5s net speech\footnote{https://github.com/CavoloFrattale/deepfake-detection-test-protocol}.

\subsection{Results}
\label{sec:results}

\begin{figure}
    \includegraphics[width=1.0\linewidth]{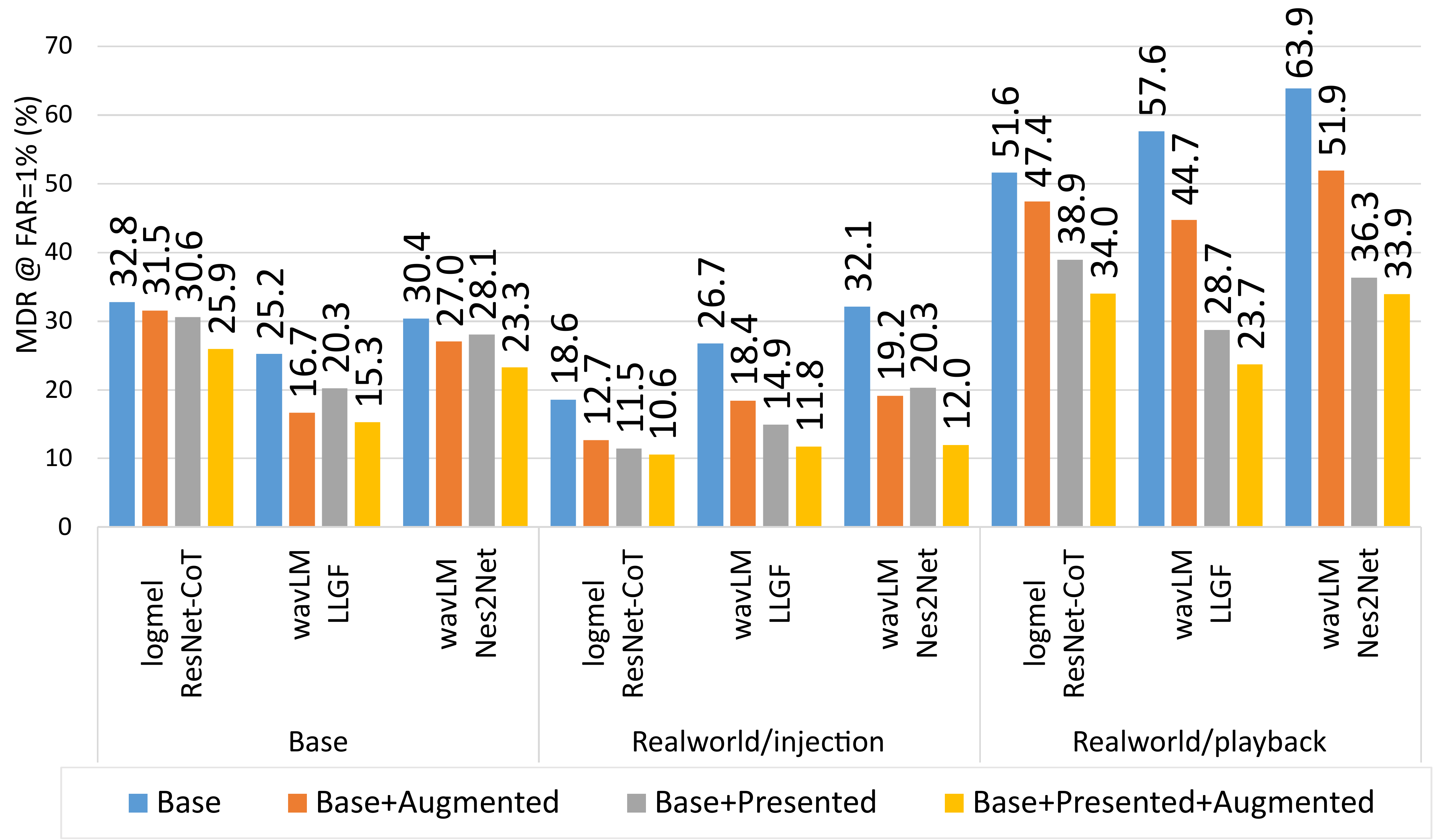}
    \caption{Plot of accuracy in terms of MDR for FAR=1\% of deepfake detection systems logmel-ResNet-CoT, WavLM-LLGF and WavLM-Nes2Net over the training conditions Base, Base+Augmented, Base+Presented, Base+Presented+Augmented (color-coded), on the testing conditions \textit{Base}, \textit{Realworld/injection}, and \textit{Realworld/playback} }
    \label{fig:overall_results}
\end{figure}

Figure~\ref{fig:overall_results} plots the MDR (y-axis, the lower the better) according to different training recipes (color-coding of bars) for each of the deepfake detection systems being evaluated (first layer of x-axis labels), across the three test conditions (second layer of x-axis labels); we see a few interesting trends and insights.

To begin, the graph demonstrates how training deepfake detection models using the datasets known to the scientific community will lead to severe decline in deepfake detection when applied to the real world (blue bars, comparing Base test condition to the Realworld ones). This degradation still occurs even though we trained systems using a larger set of datasets than previously seen in literature.

In general, using all augmentation techniques (gold bars) yields the best accuracy, especially when moving from established benchmarks (Base) to the more realistic (Realworld) scenarios.

One of our hypotheses with this study was that adding realism to training datasets and benchmarks was of paramount importance for creating deepfake detection systems that are usable in practice; today's systems as reported in scientific literature are not ready for the real world. It is remarkable, however, that adding \textit{realism} to training data (grey bars) appears to be \textbf{more} impactful than other, larger-scale augmentation techniques (orange bars); the exception to this appears in the non-realistic Base test condition, which is more aligned with the Base+Augmented training set.

Another remarkable outcome was how augmenting training data for a lightweight system with minimal computation costs (logmel-ResNet-CoT, gold bars) leads to higher accuracy than choosing larger models (WavLM-LLGF \& WavLM-Nes2Net, blue bars). Even when all systems are fully augmented (gold bars), the lighter weight logmel-ResNet-CoT model remains competitive across most conditions, even achieving the best detection rate, 89.4\% (100 - MDR) in the Realworld/Injection scenario.

Due to its much-improved performance in the Realworld playback scenario, however, the WavLM-LLGF system emerges here as the frontrunner, achieving 88.2\% deepfake detection in Realworld/Injection and 76.3\% in Realworld/playback at FAR = 1\%.

Table~\ref{tab:detailed_results} shows the detailed results for WavLM-LLGF for each dataset in the Base category. The purpose of this table is two-fold: to offer the community performance numbers on familiar benchmarks, and to establish a new baseline for comparison in future systems.

Because the majority of works in scientific literature focuses on one or very few benchmark(s), we don't intend to compare accuracy on a per-dataset basis. Rather we offer the first performance analysis across a large swath of known benchmarks and show how our methodology leads to systems that are competitive across most of them. The exceptions to this claim are the ASVspoof19 and ASVspoof21 LA results, where SOTA systems reported EER at 1\% or less, likely due to shortcut learning; the nearly identical performance of our system for ASV21LA and ASV21LA-HT confirms that we do not take advantage of silences or other shortcuts, trading off a small increase in lab benchmark error for large gains in terms of real-world robustness. It is also worth highlighting that model accuracy is consistent when comparing the Pool error to the Average, another sign of good model generalization.

\begin{table}[]
    \vspace{-6pt}
    \caption{Detailed results on the Base test condition for the selected WavLM LLGF system, in terms of Equal Error Rate (EER, \%) and Missed Detection Rate (MDR, \%) at FAR=1\% }
    \vspace{-6pt}
    \resizebox{1.0\linewidth}{!}{\begin{tabularx}{1.3\linewidth}{l|ef|ef|ef|eg}
        Train $\rightarrow$   & \multicolumn{2}{c|}{Base} & \multicolumn{2}{c|}{\shortstack[c]{Base\\+Augmented}} & \multicolumn{2}{c|}{\shortstack[c]{Base\\+Presented}} & \multicolumn{2}{c}{\shortstack[c]{Base+Presented\\+Augmented}} \\
        Test $\downarrow$ & EER & MDR & EER & MDR & EER & MDR & EER & MDR \\
        \midrule
        ASV19LA  & 6.4 & 14.3 & \textbf{4.6} & 11.2 & 5.8 & 11.1 & 5.1 & \textbf{10.5} \\
        ASV21LA  & 7.8 & 23.0 & 6.5 & 19.2 & 7.7 & 21.0 & \textbf{6.4} & \textbf{16.6} \\
        ASV21LA-HT & 8.2 & 21.5 & 7.4 & 22.8 & 7.5 & 19.3 & \textbf{6.5} & \textbf{17.5} \\
        ASV21DF  & 4.7 & 10.7 & \textbf{3.1} & \textbf{6.6} & 4.0 & 9.7 & 3.3 & 6.7 \\
        ASV5 w/o Enc. & 8.2 & 46.7 & 5.5 & 48.4 & 4.4 & \textbf{28.4} & \textbf{3.8} & 33.8 \\
        In-the-wild & 6.9 & 28.0 & \textbf{2.1} & \textbf{3.3} & 4.7 & 20.3 & 2.8 & 7.3 \\
        SpoofCeleb  & 7.5 & 23.0 & \textbf{5.2} & \textbf{11.8} & 6.5 & 22.2 & 5.8 & 17.8 \\
        \midrule
        Pool & 7.1 & 25.2 & 5.0 & 16.7 & 6.0 & 20.3 & \textbf{4.8} & \textbf{15.3} \\
        Average & 7.1 & 24.1 & 4.9 & 17.5 & 5.8 & 19.0 & \textbf{4.8} & \textbf{15.7} \\
        \bottomrule        
    \end{tabularx}}
    \label{tab:detailed_results}
\end{table}

\section{Conclusions}
\label{sec:conclusions}
We proposed a methodology for conducting research on and developing deepfake detection technology. We sought to demonstrate how this new methodology would largely improve deepfake detection systems' generalization across a broad array of attacks, and lead to systems that are more effective in real world applications.

Our results successfully reinforced our hypotheses, with a few unexpected insights. Namely, we observed that adding realism (in our case, by incorporating \textit{presentation} methods) to the training data was more important than augmenting it using other techniques (but ideally you should do both), especially (and perhaps unsurprisingly) when it applied to the Realworld dataset.

We also observed that with full data augmentation the lightweight logmel-ResNet-CoT system was highly competitive with the much larger WavLM-based systems. Given the excessive costs (and likely need for GPU-based inference) of the larger models, this result suggests that research groups should first invest resources towards improved data collection before exploring the added benefits of larger pre-trained models.

WavLM-LLGF was the strongest overall system, but its 76.3\% detection rate in the Realworld/playback scenario should be further improved. Including a specialized playback detection model \cite{wang20_asvspoof,asvspoof2021} would likely improve accuracy without loss of generality.

Researchers developing deepfake detection technology should continue to be mindful of the assumptions around existing public datasets and benchmarks, and make adjustments accordingly. A complete understanding of the real-world scenario is also paramount. Deepfake technologies have not only evolved at a rapid pace in recent years, but they are increasingly being deployed in the real world to nefarious ends. It is therefore our collective responsibility to also evolve our own approach towards developing deepfake detection systems to ensure that the public is well-defended.

\vspace{-10pt}
\section{Compliance with ethical standards}
This research study was conducted retrospectively using human subject data made available in open access by ASVspoof and LDC, or restricted to internal usage at Microsoft. Ethical approval was not required as confirmed by the license attached with the open access data and Microsoft policy on data protection and privacy in the case of the internal dataset.

\bibliographystyle{IEEEtran}
\bibliography{refs}

\end{document}